\documentclass[12pt]{article}
\usepackage{amsmath}
\usepackage{bm}
\usepackage{color}
\usepackage{caption}
\usepackage{subcaption}
\usepackage{graphicx}
\usepackage{braket}
\begin{document}
\title{Energy levels estimation on a quantum computer by evolution of a physical quantity}

\author{
Kh. P. Gnatenko$^{1,2}$\footnote{khrystyna.gnatenko@gmail.com}, H. P. Laba$^3$\footnote{hanna.laba@polynet.lviv.ua},
V. M. Tkachuk$^1$\footnote{voltkachuk@gmail.com}\\
$^1$Professor Ivan Vakarchuk Department for Theoretical Physics,\\
Ivan Franko National University of Lviv,\\
12, Drahomanov St., Lviv, 79005, Ukraine.\\
$^2$ SoftServe Inc., Austin, TX, USA.\\
$^3$Department of Applied Physics and Nanomaterials Science, \\
Lviv Polytechnic National University,\\
5 Ustiyanovych St., 79013 Lviv, Ukraine.}

\maketitle

\begin{abstract}
We show that the time dependence of mean value of a physical quantity is related with the transition energies of a quantum system. In the case when the operator of a physical quantity anticommutes with the Hamiltonian of a system, studies of the evolution of its mean value allow determining the energy levels of the system. On the basis of the result, we propose a method for determining energy levels of physical systems on a quantum computer. The method opens a possibility to achieve quantum supremacy in solving the problem of finding minimal or maximal energy of Ising model with spatially anisotropic interaction using multi-qubit quantum computers. We apply the method for spin systems (spin in magnetic field, spin chain, Ising model on squared lattice) and realize it on IBM's quantum computers.


\end{abstract}

\section{Introduction}
Calculation of energy spectrum of a
Hamiltonian is an  important  problem in quantum mechanics. Recent development of quantum computers recognizes considering them as a powerful tool for solving this problem.

One of the algorithms allowing determination
of eigenvalues for Hamiltonian is the quantum phase estimation (QPE) algorithm \cite{Abrams97,Abrams99,Kitaev97,Dob,Paesani,Parker20}. It was
originally proposed by Kitaev, Lloyd and Abrams \cite{Abrams97,Abrams99,Kitaev97}.
This algorithm is based on finding eigenvalue $\lambda=e^{i\phi}$ or phase $\phi$  of a unitary operator. In the case when the unitary operator is the operator of evolution of a quantum system the phase $\phi$ is related with eigenvalues of the Hamiltonian. Short review on this problem can be found in \cite{Cruz20}. In \cite{Russo} the method for estimation of the transition energies on the basis of robust phase estimation algorithm  was proposed.

Also, hybrid classical-quantum algorithms that allows to examine energy levels are known. Among them are  quantum approximate optimization algorithm  (it recognizes to find ground state energy and is used to solve optimization problems \cite{Farhi, Farhi1,Moll,Fuchs}), variational quantum eigensolver (it recognizes to obtain the transition energies \cite{Peruzzo,Mc,Malley,Parrish}).

In \cite{Somma19} the authors presented an efficient method for estimating the eigenvalues of a Hamiltonian from the time dependence of expectation values of the evolution operator. Originally this idea  was suggested in \cite{Somma02}.  In \cite{Liu} qubit efficient circuit architecture was addopted for the variational quantum eigensolver and qubit efficient scheme to study ground-state properties of quantum many-body systems on a quantum computer was introduced. In  \cite{Motta} quantum algorithms (quantum Lanczos, quantum analogue of the minimally entangled typical thermal states,quantum analogue of the minimally entangled typical thermal states)  that gives a possibility to detect ground, excited and thermal states on a quantum computer were described.

In this paper we show that studies of time dependence of mean value of a physical quantity allow to extract transition energies of a quantum system. In the case when the operator of the physical quantity anticommutes with the Hamiltonian such a studies give a possibility to determine the energy levels of a system. On the basis of the results we propose method for detection of energy levels of physical systems on a quantum computer. Using the method, energy levels of spin systems (spin in magnetic field, spin chain, Ising model on squared lattice) are found performing calculations on IBM's quantum computers

The paper is organized as follows. In Section 2 we propose a method to estimate transition energies on the basis of studies of evolution of the mean value of physical quantity. In Section 3 we show that in particular case when the operator of a physical quantity anticommutes with the Hamiltonian of a physical system studies of evolution of the mean values of the operator gives a possibility to detect the energy levels of the physical system. In  Section 4 we present results for energy levels of spin systems obtained on the basis of calculations on the IBM's quantum computers. Conclusions are presented in Section 5.

\section{Evolution of mean value of physical quantity and transition energies of a physical system}

Let us consider a physical system with Hamiltonian $H$.
 The evolution of a state vector of the system in time can be written as follows
\begin{eqnarray}
|\psi(t)\rangle=e^{-iH t/\hbar}|\psi_0\rangle=\sum_{i}c_ie^{-iE_i t/\hbar}|E_i\rangle,
\end{eqnarray}
where we use notation $|\psi_0\rangle$ for the state of the system at the
initial moment of  time $t=0$  and expand it over the eigenstates  of Hamiltonian $|E_i\rangle$  namely,
\begin{eqnarray}\label{psi0E}
|\psi_0\rangle=\sum_{i}c_i|E_i\rangle.
\end{eqnarray}
$E_i$ are energy levels of the system.

Let us consider a physical quantity represented by operator $\hat A$
Then evolution of mean value of the quantity reads
\begin{eqnarray}\label{At}
A(t)=\langle \psi(t)| \hat A|\psi(t)\rangle
=\sum_i\sum_jc^*_ic_je^{i\omega_{ij}t}A_{ij}=\sum_i\sum_jg_{ij}e^{i\omega_{ij}t},
\end{eqnarray}
where $\omega_{ij}=(E_i-E_j)/\hbar$ is frequency of transition between energy levels
$E_i$ and $E_j$, $A_{ij}=\langle E_i|\hat A|E_j\rangle$ is matrix element of operator
$\hat A$ and $g_{ij}=c^*_ic_jA_{ij}$ is hermitian matrix

The goal of this paper is to extract from function $A(t)$ the frequencies
$\omega_{ij}$. For this purpose we consider the following transformation of expectation value
\begin{eqnarray}\label{Aomega}
A(\omega)={1\over 2\pi}\int_{-\infty}^{\infty}dt A(t)e^{-i\omega t}.
\end{eqnarray}

Substituting (\ref{At}) into (\ref{Aomega}) we find
\begin{eqnarray}\label{Adelta}
A(\omega)=\sum_i\sum_jg_{ij}\delta(\omega-\omega_{ij}).
\end{eqnarray}
Thus function $A(\omega)$ has $\delta$ - peaks at $\omega=\omega_{ij}$. It allows
knowing  $A(\omega)$ to find the frequencies of transitions $\omega_{ij}$.
Note that $A^*(\omega)=A(-\omega)$ and in general contains real and imaginary parts
\begin{eqnarray}
A(\omega)=A_1(\omega)+iA_2(\omega),
\end{eqnarray}
where
\begin{eqnarray}
A_1(\omega)=\int_0^{\infty}dt A(t)\cos(\omega t),\\
A_2(\omega)=\int_0^{\infty}dt A(t)\sin(\omega t).
\end{eqnarray}

In order to apply this result for finding the frequencies of transitions on a quantum computer we have to take into account that on a quantum computer we can find mean values of physical quantity $A(t)$ at some fixed moments of time. Thus we write $t=\tau n$, were $n=-N, -N+1,...N-1, N$ and $\tau$ is some fixed time interval. Then for $A(\omega)$ we have
\begin{eqnarray} \nonumber
A(\omega)=\sum_i\sum_jg_{ij}{1\over 2\pi}\sum_{n=-N}^N \tau e^{-i(\omega-\omega_{ij})\tau n}=\\
\sum_i\sum_jg_{ij}{1\over 2\pi}\tau\left( 1 + 2\cos((\omega-\omega_{ij})(T+\tau)/2)  {\sin((\omega-\omega_{ij})T/2)\over\sin((\omega-\omega_{ij})\tau/2)}\right),
\end{eqnarray}
where $T=N\tau$. At $\tau\to 0$ and fixed $T$ we find
\begin{eqnarray} \label{AomegaT}
A(\omega)=\sum_i\sum_jg_{ij}{1\over\pi}{\sin((\omega-\omega_{ij})T)\over(\omega-\omega_{ij})}.
\end{eqnarray}
Each term in this expression takes maximal value at $\omega=\omega_{ij}$  which is
$Tg_{ij}/\pi$.

Note that (\ref{AomegaT}) corresponds to interval of integration in (\ref{Aomega}) from $-T$ to $T$.
When in additional $T\to \infty$ then (\ref{AomegaT}) tends to (\ref{Adelta}).

In general the idea to use these results for finding frequencies of transition between different energy levels is the following. We study evolution of quantum system and find time dependence of some physical quantity $A(t)$  on a quantum computer. Then we find Fourier transformation $A(\omega)$ of this function. Maximal values of $A(\omega)$ allow to detect the frequencies of transitions.

\section{Detecting of energy levels of a physical system by evolution of mean value}

In this section we show that in the particular case studying evolution of the mean value of operator $\hat A$ one can determine the directly energy levels of a physical system.
 Let us consider operator which anticommutes with Hamiltonian of a system
\begin{eqnarray}
\{\hat A, H\}=0.
\end{eqnarray}
If such operator exists the energy spectrum of a system is symmetric with respect to  $E\to-E$.
In other worlds if $E$ is the energy level of a system, $-E$ is the energy level too. In order to show this let us consider stationary Schr\:odinger equation
\begin{eqnarray}
H|\psi\rangle=E|\psi\rangle.
\end{eqnarray}
Applying operator $\hat A$ anticommuting with $H$ to left and right hand side  of the equation we have
\begin{eqnarray}
H\hat A|\psi\rangle=-E\hat A|\psi\rangle.
\end{eqnarray}
Thus $A|\psi\rangle$ is eigenstate with energy $-E$. Of course here we assume that
state $A|\psi\rangle$ exists.
In this case for mean value of the operator $\hat{A}$ we find
\begin{eqnarray}
A(t)=\langle\psi_0|e^{iHt/\hbar} \hat A e^{-iHt/\hbar}|\psi_0\rangle=\\
\langle\psi_0|e^{2iHt/\hbar} \hat A |\psi_0\rangle=
\langle\psi_0|\hat Ae^{-2iHt/\hbar}  |\psi_0\rangle,
\end{eqnarray}
where we  taking into account that  $\hat A e^{-iHt/\hbar}= e^{iHt/\hbar}\hat A$.

Representing initial state in the form (\ref{psi0E}) the mean value reads
\begin{eqnarray}
A(t)=\sum_j g_je^{-i2\omega_jt},
\end{eqnarray}
where $g_j=\sum_ic_i^*A_{ij}c_j$.
The mean value $A(t)=A^*(t)$ is real, so we can also write
\begin{eqnarray}
A(t)=\sum_j g^*_je^{i2\omega_jt}.
\end{eqnarray}
Then $A(\omega)$
has a sharp pics at $\omega=2\omega_i$
\begin{eqnarray}
A(\omega)=\sum_j g^*_j \delta(\omega-2\omega_i).\label{aww}
\end{eqnarray}
So, analysis of $A(\omega)$ allow to determine the energy spectrum of a physical system.
In the next Section we apply this method for spin systems that are one of the most suitable systems for modeling them on a quantum computer.

\section{Detecting energy levels of spin systems on IBM's quantum computer}
\subsection{Spin in magnetic field}
As the first example we consider a spin in magnetic field directed along $z$-axis
with Hamiltonian
\begin{eqnarray}
H={\hbar\omega_0}\sigma^{z}.\label{hams1}
\end{eqnarray}
Let us apply the method proposed in the previous section and find the energy levels of the system using a quantum computer.

It is easy to write operator that anticommutes with the Hamiltonian (\ref{hams1}). For example one can choose   $\hat{A}=\sigma_x$ ($\{\sigma_x,{\hbar\omega_0}\sigma^{z}\}=0$) and study its mean value.  We consider the initial state as
\begin{eqnarray}\label{inits1}
|\psi_0\rangle={1\over\sqrt 2}(|0\rangle+|1\rangle)=|+\rangle.
\end{eqnarray}
The state (\ref{inits1}) is the eigenstate of operator $\sigma^x$ with eigenvalue $1$ and is not the eigenstate of Hamiltonian (\ref{hams1}).
So, the mean value of $\hat{A}$ reads
\begin{eqnarray}
A(t)=\langle\sigma^x(t)\rangle=\langle +
|e^{i\omega_0\sigma^zt}\sigma^xe^{-i\omega_0\sigma^zt}|+\rangle=\nonumber\\
=\langle +|e^{2i\omega_0\sigma^zt}\sigma^x|+\rangle=
{1\over 2}\left(e^{2i\omega_0t}+ e^{-2i\omega_0t}\right).\label{at0}
\end{eqnarray}
Then function
\begin{eqnarray}
A(\omega)={1\over 2}(\delta(\omega-2\omega_0)+\delta(\omega+2\omega_0))
\end{eqnarray}
has peaks at $\omega=\pm 2\omega_0$ that are double eigenvalues of Hamiltonian (\ref{hams1}) in the units of $\hbar$.

\begin{figure}[!!h]
\begin{center}
\includegraphics[scale=0.65, angle=0.0, clip]{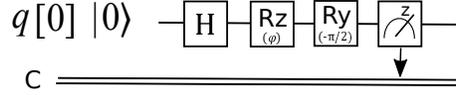}
\end{center}
\caption{Quantum protocol for studies of mean value evolution  $\langle\sigma^x(t)\rangle$ in the case of spin in the magnetic field.}
		\label{fig:1}
\end{figure}

To study evolution of the mean value $\langle\sigma^x(t)\rangle$ given by (\ref{at0}) on a quantum computer we consider quantum protocol Fig. \ref{fig:1}.
 Applying Hadamar gate $H$ and Z-rotation gate   $RZ(\phi)$ one prepares the state $\exp(-i\phi \sigma^z/2)\ket +$. To measure the mean value of the operator $\sigma^x$ in the state $\exp(-i\phi \sigma^z/2)\ket +$, we take into account that the operator $\sigma^x$ can be represented as $\sigma^x=\exp(-i\pi\sigma^y/4)\sigma^z\exp(i\pi\sigma^y/4)$. Therefore to find $\langle\sigma^x\rangle$ the Y-rotation gate $RY(\pi/2)$ is applied before measurement in the standard basis  (see Fig. \ref{fig:1}).  The value $\langle\sigma^x\rangle$ can be calculated using  the results of measurement as
\begin{eqnarray}
\langle\sigma^x\rangle=\langle +
|e^{i\phi\sigma^z/2}\sigma^xe^{-i\phi\sigma^z/2}|+\rangle=\bra{\tilde\psi}\sigma^z\ket{\tilde\psi}=
\vert \langle \tilde\psi\vert 0 \rangle \vert^2-\vert \langle \tilde\psi\vert 1 \rangle \vert^2,\label{ms}
\end{eqnarray}
where
\begin{eqnarray}
\ket{\tilde\psi}=e^{i \pi\sigma^y/4}e^{-i\phi\sigma^z/2}|+\rangle.
\end{eqnarray}
Applying quantum protocol  Fig. \ref{fig:1} with different values of parameter $\phi=2\hbar\omega_0t$ we measure dependence of the mean value of $\langle\sigma^x\rangle$ on time.
Such studies were done on  IBM's quantum computer $\textrm{ibmq\_manila}$. The structure of the quantum device is presented on Fig. 2.

\begin{figure}[!!h]
\begin{center}
\includegraphics[scale=0.65, angle=0.0, clip]{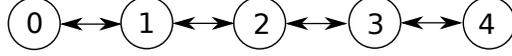}
\end{center}
\caption{Structure of IBM's quantum computer $\textrm{ibmq\_manila}$. Arrows link qubits between which the controlled NOT gate can be directly applied.}
		\label{t1}
\end{figure}

\begin{figure}[!!h]
\begin{center}
\includegraphics[scale=0.3, angle=0.0, clip]{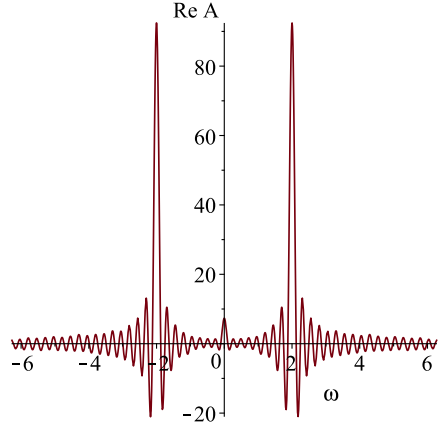}
\end{center}
\caption{Result of detecting of the energy levels of a spin in magnetic field using studies of evolution of mean value $\langle \sigma^x\rangle$ on IBM's quantum computer $\textrm{ibmq\_manila}$. The real part of $A(\omega)$ has peaks at $\omega=\pm 2$ that correspond to energies $E=\pm\hbar\omega_0$.}
		\label{fig:5}
\end{figure}

 We realized quantum protocol Fig. \ref{fig:1} on qubit $q[0]$ of $\textrm{ibmq\_manila}$  for parameter $\phi/2=\hbar\omega_0t$ changing from $-8\pi$ to $8\pi$ with the step $\pi/12$. On the basis of the obtained results taking into account (\ref{ms}) we find mean values  $\langle\sigma^x(t)\rangle$ for $t=n\tau$, $\tau=\pi/12\omega_0$, $n=-N,-N+1,...N-1,N$, N=96 and calculate
\begin{eqnarray}
A(\omega)=\sum^{N}_{n=-N}\langle\sigma^x(n\tau)\rangle e^{i \omega n \tau}\label{aw0}
\end{eqnarray}
The results of calculations are presented on Fig. \ref{fig:5}. For convenience we put $\omega_0=1$. From Fig. \ref{fig:5} we see that the real part of $A(\omega)$  has peaks at $\omega=\pm 2$ that according to (\ref{aww}) correspond to energies $E=\pm\hbar\omega_0$.

\subsection{Spin chain}
\begin{figure}[!!h]
\begin{center}
\includegraphics[scale=0.65, angle=0.0, clip]{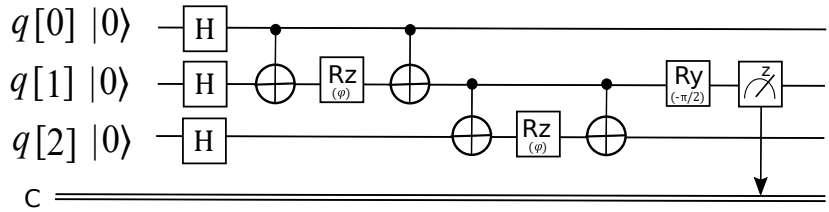}
\end{center}
\caption{Quantum protocol for studies of mean value evolution  $\langle\sigma^x(t)\rangle$ in the case of spin chain, $\phi=2Jt/\hbar$}
		\label{fig:6}
\end{figure}

As the second example we consider a chain of three spins with the Ising interaction described by the following Hamiltonian
\begin{align}
H=J\sigma^z_0\sigma^z_1+J\sigma^z_1\sigma^z_2,\label{hams2}
\end{align}
here $J$ is the interaction coupling.
In this case we chose operator $\hat A$ to be $\sigma^x_1$. Note that $\{\sigma^x_1,H\}=0$.
Quantum protocol for studies of  $\langle\sigma_1^x(t)\rangle$ at different moments of time is presented on Fig.  \ref{fig:6}.

In the quantum protocol  we take into account that operator $\exp(-iJ\sigma^z_i\sigma^z_j/\hbar)$ with exactness to total
phase factor can be rewritten as $CX_{ij}RZ_j(2Jt/\hbar)CX_{ij}$, here $CX_{ij}$ is the controlled-NOT
gate that acts on qubit  $q[i]$ as a control and on qubit $q[j]$ as a target, Z-rotation gate $RZ_j(2Jt /\hbar)$ acts on qubit  $q[i]$.  So, with exactness to total phase factor the state $e^{-iHt/\hbar}|+\rangle$ is prepared  in the result of action of gates $CX_{01}RZ_1(2Jt /\hbar)CX_{01}CX_{12}RZ_2(2Jt/\hbar)CX_{12}H_0H_1H_2$ on  the state $\ket{000}$ (see Fig.  (\ref{fig:6})).  Then to detect the mean value $\langle\sigma^x_1(t)\rangle$ similarly as in the previous subsection we implement rotation of the state of qubit $q[1]$ around the $x$ axis by $\pi/2$ and perform measurement of the state in the standard basis (see Fig.  (\ref{fig:6})).

\begin{figure}[!!h]
\begin{center}
\includegraphics[scale=0.3, angle=0.0, clip]{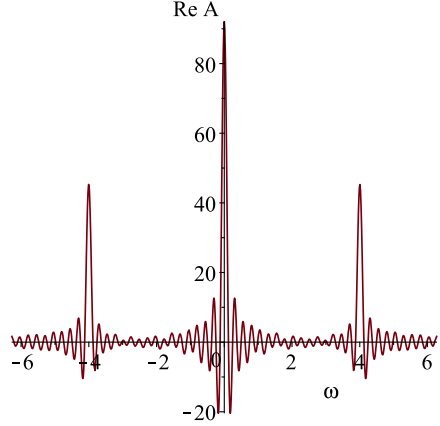}
\end{center}
\caption{Result of detecting of the energy levels of the spin chain  (\ref{hams2}) by studies of evolution of mean value $\langle \sigma_1^x\rangle$ on IBM's quantum computer $\textrm{ibmq\_manila}$. Peaks of $\textrm{Re}A(\omega)$ at $\omega=0$ and $\omega=\pm 4$ correspond to energies $E=0$, $E=\pm2J$.}
		\label{fig:7}
\end{figure}

 Quantum protocol Fig. \ref{fig:6} was realized on qubits  $q[0]$, $q[1]$, $q[2]$ of $\textrm{ibmq\_manila}$. Note that the quantum device has the chain structure (see Fig. \ref{fig:1}).  So,  the CNOT gates in  Fig.  \ref{fig:6} can be applied directly to the respective qubits. Changing $\phi/2=Jt/\hbar$  from $-8\pi$ to $8\pi$ with the step $\pi/12$, we detect the mean values $\langle\sigma^x(t)\rangle$  for $t=n\tau$, $\tau=\hbar\pi/J12$, $n=-N,-N+1,...N-1,N$, N=96 and calculate $A(\omega)$ (\ref{aw0}). The results of calculations are presented in Fig. \ref{fig:7}. In Fig. \ref{fig:7} for convenience we put $\hbar/J=1$. The real part of $A(\omega)$ has peaks at $\omega=0$ and $\omega=\pm 4$ that correspond to the energy levels $E=0$, $E=\pm2J$, respectively. 

\subsection{Ising model on squared latice}
Let us consider the Ising model
\begin{eqnarray}
H={1\over 2}\sum_{i,j}J_{ij}\sigma^z_i\sigma^z_j.\label{IZ}
\end{eqnarray}
where $J_{ij}$ are the interaction coupling.

We study square-lattices with  isotropic and spatially anisotropic Ising interactions  described, respectively,  by the following Hamiltonians
\begin{align}
H_1=J\sigma^z_0\sigma^z_1+J\sigma^z_1\sigma^z_2+J\sigma^z_2\sigma^z_3+J\sigma^z_3\sigma^z_0,\label{hams3}\\
H_2=-J\sigma^z_0\sigma^z_1+J\sigma^z_1\sigma^z_2+J\sigma^z_2\sigma^z_3+J\sigma^z_3\sigma^z_0, \label{hams4}
\end{align}
and detect the corresponding energy levels on IBM's quantum computer $\textrm{ibmq\_manila}$. 

\begin{figure}[!!h]
\begin{center}
{\includegraphics[scale=0.43, angle=0.0, clip]{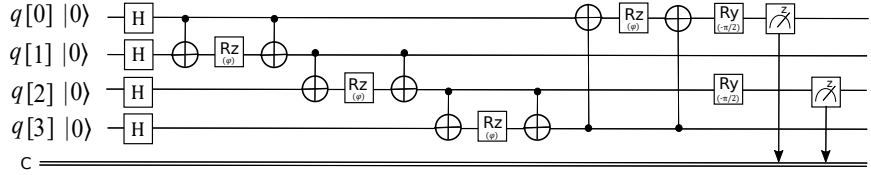}}
\end{center}
\caption{Quantum protocol for studies of mean value evolution  $\langle\sigma^x_0(t)\sigma^x_2(t)\rangle$ in the case of the Ising model on squared latice with (\ref{hams3}), $\phi=2Jt/\hbar$. }
		\label{fig:8}
\end{figure}

\begin{figure}[!!h]
\begin{center}
\subcaptionbox{\label{ff2}}{\includegraphics[scale=0.3, angle=0.0, clip]{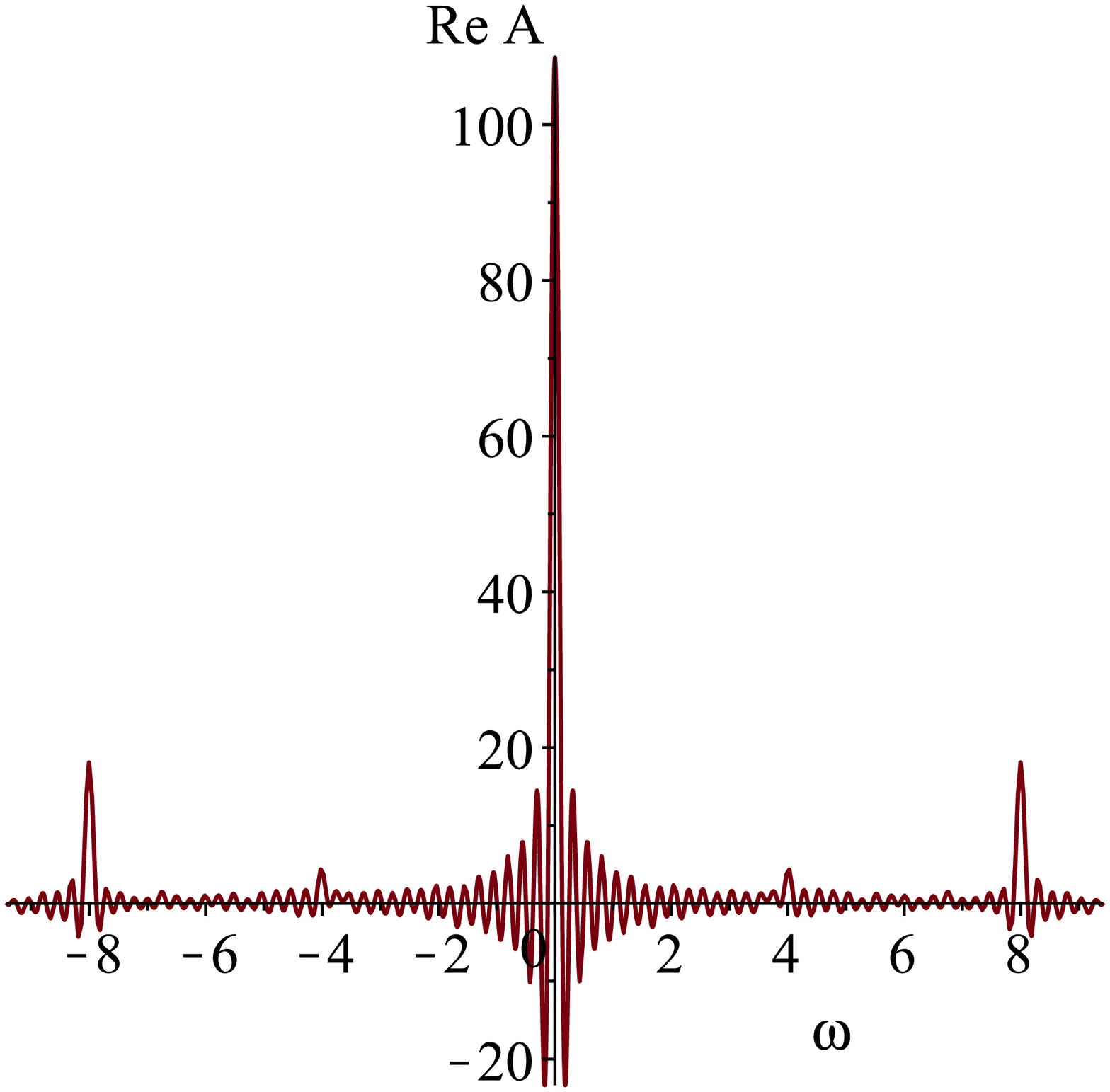}}
\subcaptionbox{\label{ff2}}{\includegraphics[scale=0.3, angle=0.0, clip]{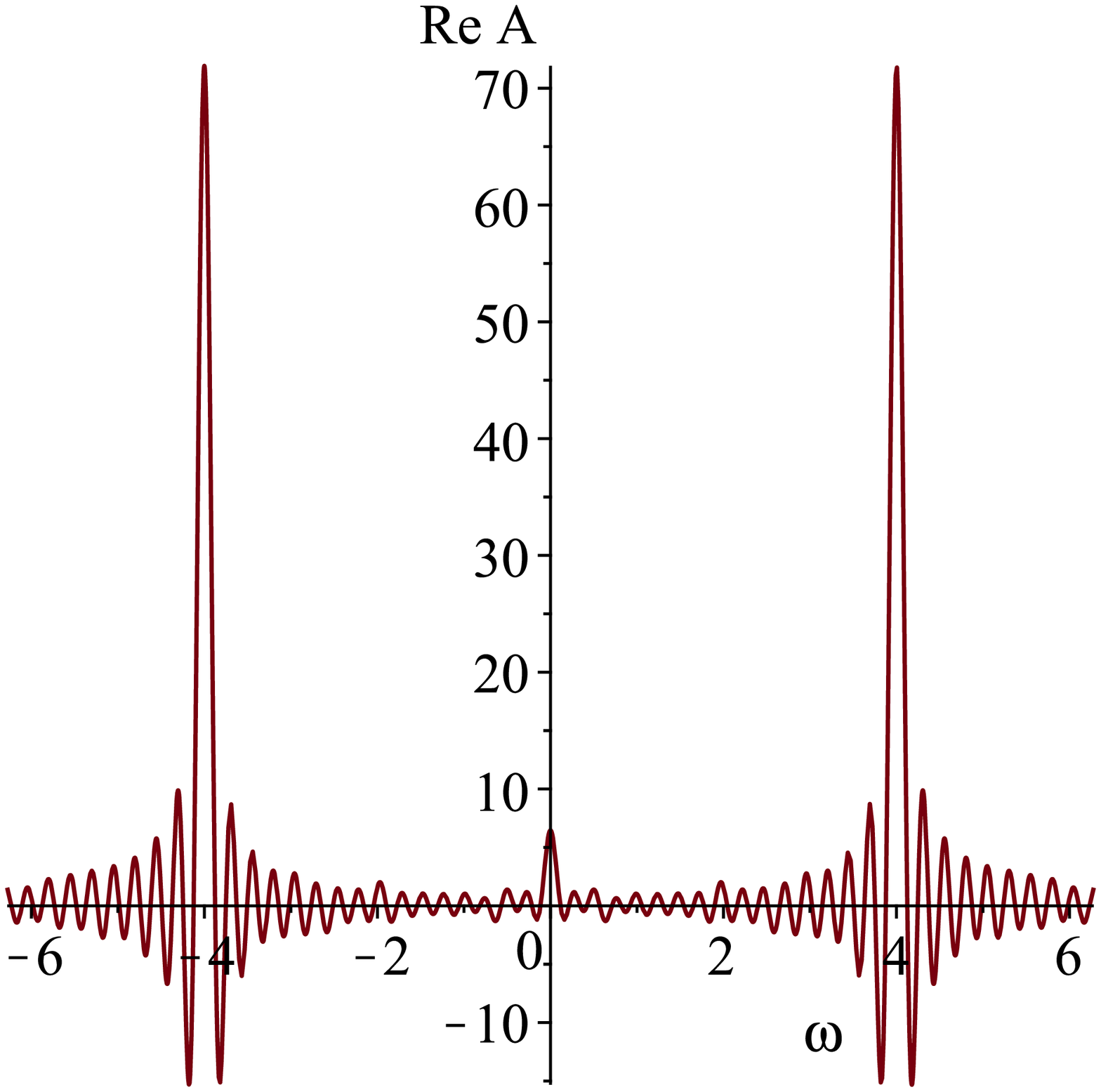}}
\end{center}
\caption{Results of detecting of the energy levels of  Ising model on squared latice with isotropic (\ref{hams3}) (a) and  spatially anisotropic Ising interactions (\ref{hams4}) (b)    by studies of evolution of mean value $\langle \sigma_0^x\sigma_2^x\rangle$ on IBM's quantum computer $\textrm{ibmq\_manila}$. The peaks of $\textrm{Re}A(\omega)$ (a) at $\omega=0$ and $\omega=\pm 8$ correspond to energies $E=0$, $E=\pm4J$. The peaks of $\textrm{Re}A(\omega)$ (b) at $\omega=\pm 4$ correspond to energies  $E=\pm2J$.}
		\label{fig:9}
\end{figure}

As the initial states we chose $|\psi_0\rangle=|++++\rangle$. To detect the energy levels of the corresponding systems we examine evolution of the operator $\hat A=\sigma^x_0\sigma^x_2$, ($\{\sigma^x_0\sigma^x_2,H_1\}=\{\sigma^x_0\sigma^x_2,H_2\}=0$).
Quantum protocol for studies of $\braket {A(t)}$ is presented on Fig. \ref{fig:8}.
To examine the case of spatially anisotropic Ising interaction (\ref{hams4}) we change the sign of parameter $\phi$ in $RZ$ gate acting on the qubit $q[1]$.
The results for discrete Fourier transformation $A(\omega)$  are presented  on Fig. \ref{fig:9}. Function $\textrm{Re}A(\omega)$ (Fig.  \ref{fig:9} (a)) has peaks at $\omega=0$ and $\omega=\pm 8$ that correspond to energies $E=0$, $E=\pm4J$ of the Ising model on squared latice (\ref{hams4}).  In the case of  Ising model with spatially anisotropic Ising interaction the function  $\textrm{Re}A(\omega)$  has peaks  at $\omega=\pm 4$ (Fig.  \ref{fig:9} (b))  corresponding to the energies  $E=\pm2J$.

\begin{figure}[!!h]
\begin{center}
\includegraphics[scale=0.65, angle=0.0, clip]{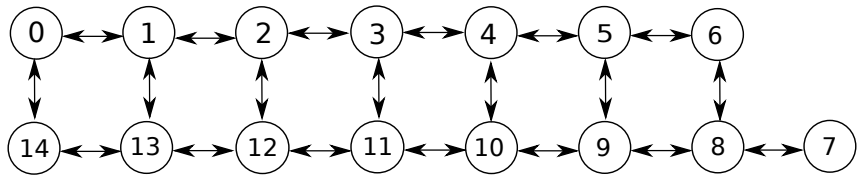}
\end{center}
\caption{Structure of IBM's quantum computer $\textrm{ibmq\_melbourne}$. Arrows link qubits between which the controlled NOT gate can be directly applied.}
		\label{fig:10}
\end{figure}

Finally, we also detect energy levels of a Ising model on squared latice in the case of six spins (qubits). The calculations were made on 15-qubit quantum device $\textrm{ibmq\_melbourne}$ with structure given in Fig. \ref{fig:10} and on the quantum simulator $\textrm{ibmq\_
qasm\_simulator}$.

\begin{figure}[!!h]
\begin{center}
\includegraphics[scale=0.35, angle=0.0, clip]{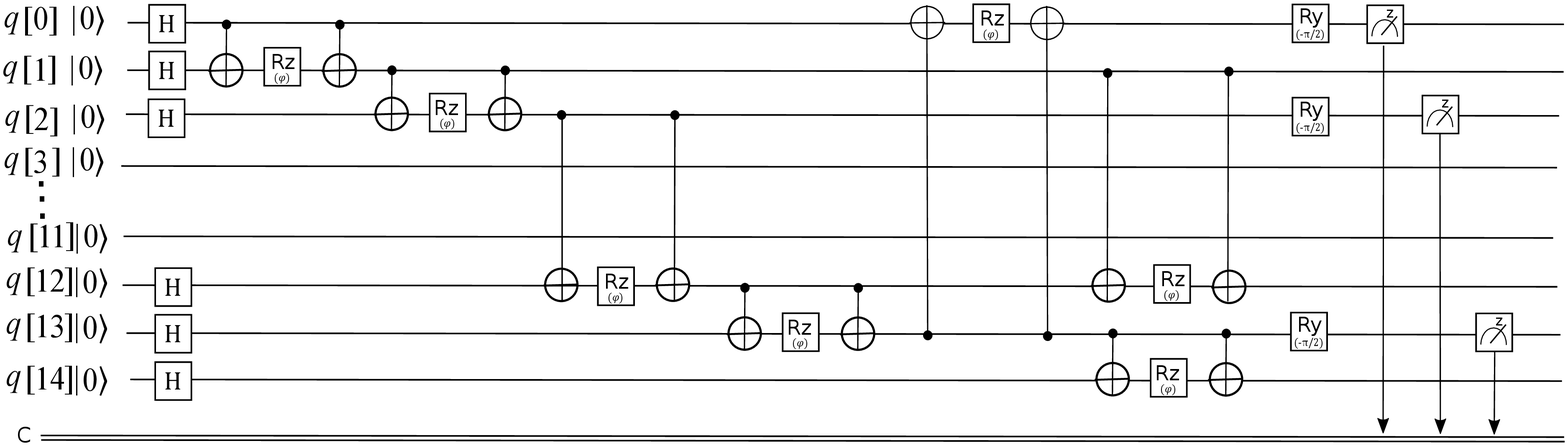}
\end{center}
\caption{Quantum protocol for studies of mean value evolution  $\langle\sigma^x_0(t)\sigma^x_2(t)\sigma^x_{13}(t)\rangle$ in the case of the Ising model on squared latice with (\ref{hams5}), $\phi=2Jt/\hbar$. }
		\label{fig:11}
\end{figure}

Taking into account the structure of the quantum computer $\textrm{ibmq\_melbourne}$ we study a system described by the following Hamiltonian
\begin{align}
H=J\sigma^z_0\sigma^z_1+J\sigma^z_1\sigma^z_2+J\sigma^z_2\sigma^z_{12}+J\sigma^z_{13}\sigma^z_{12}+J\sigma^z_{13}\sigma^z_{14}+J\sigma^z_1\sigma^z_{13}+J\sigma^z_0\sigma^z_{14},\label{hams5}
\end{align}
We choose the initial state to be $\ket{++++++}$ and study the evolution of mean value of operator   $\hat A=\sigma^x_0\sigma^x_2\sigma^x_{13}$. Note that $\hat A$ anticommutes with Hamiltonian (\ref{hams5}).  For this purpose quantum protocol Fig. \ref{fig:11} was implemented for different values of $\phi=2Jt/\hbar$ on qubits $q[0]$, $q[1]$, $q[2]$, $q[12]$, $q[13]$, $q[14]$  of $\textrm{ibmq\_melbourne}$.

\begin{figure}[!!h]
\begin{center}
\subcaptionbox{\label{ff1}}{\includegraphics[scale=0.3, angle=0.0, clip]{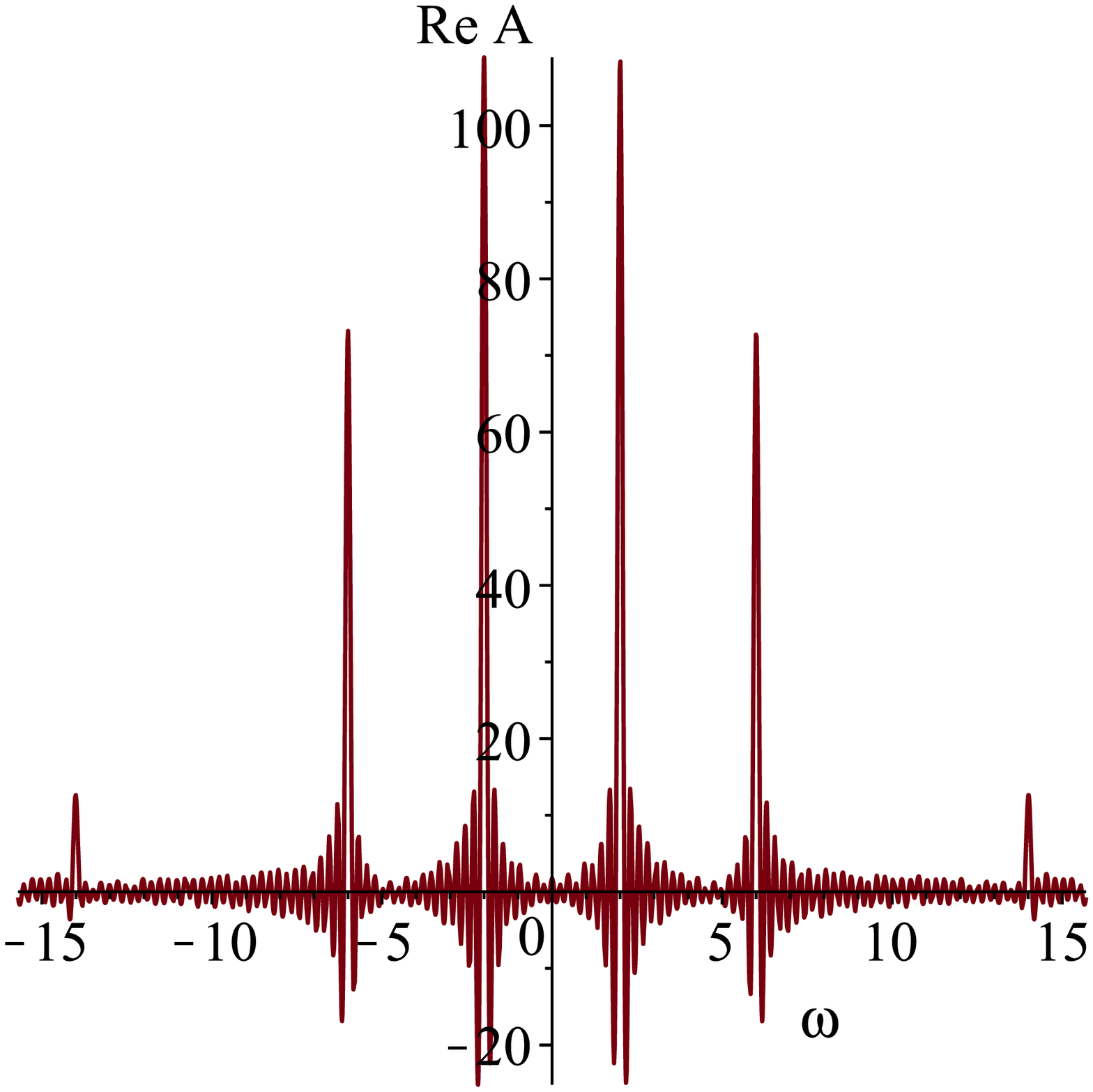}}
\hspace{1cm}
\subcaptionbox{\label{ff3}}{\includegraphics[scale=0.3, angle=0.0, clip]{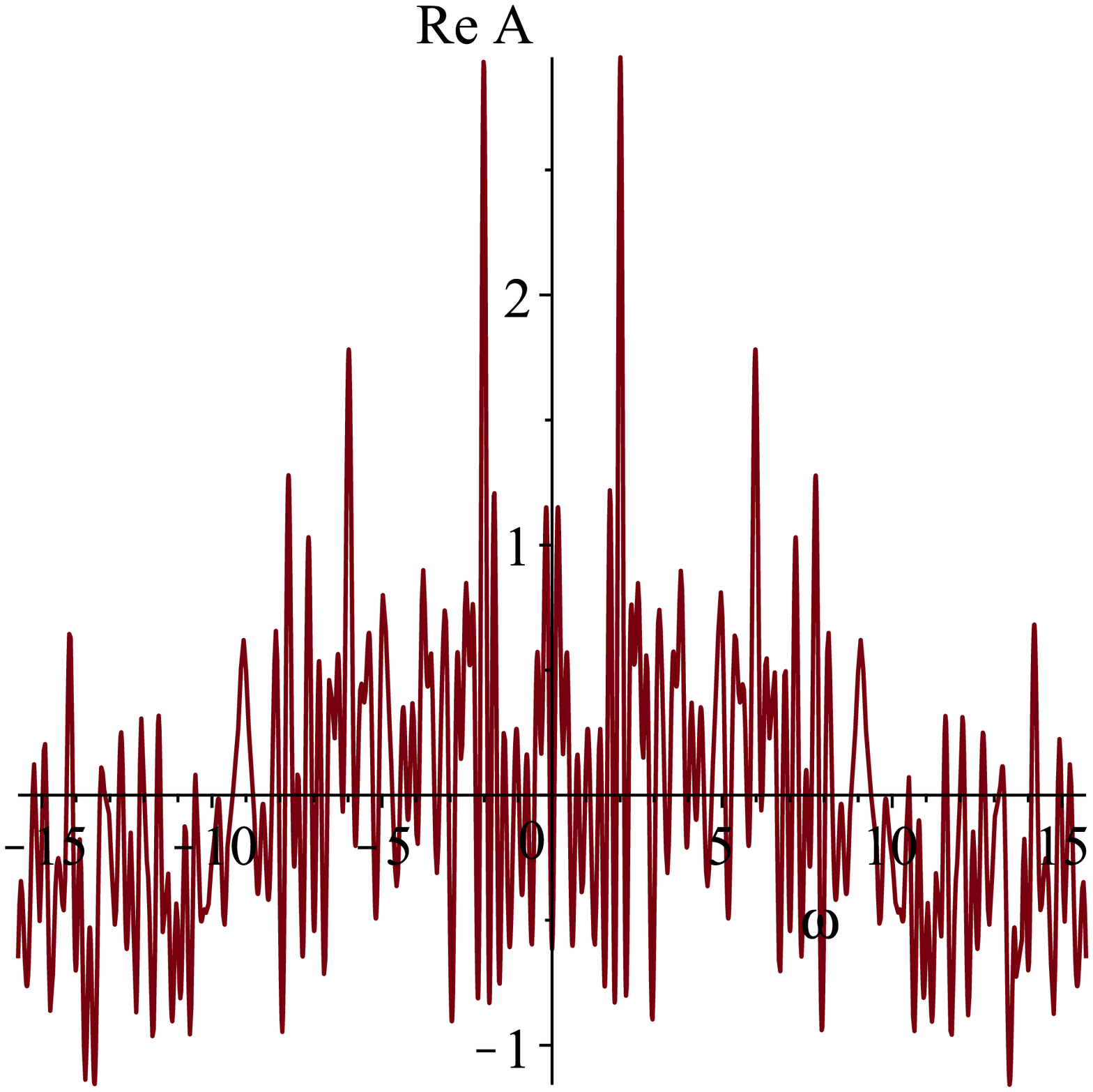}}
\end{center}
\caption{Results of detecting of the energy levels of  Ising model on squared latice (\ref{hams5}) on $\textrm{ibmq\_qasm\_simulator}$ (a) and   $\textrm{ibmq\_melbourne}$ (b)    by studies of evolution of mean value $\langle \sigma_0^x\sigma_2^x\sigma_{13}^x\rangle$. The peaks of $\textrm{Re}A(\omega)$ at $\omega=0$, $\omega=\pm 6$,  $\omega=\pm 14$ correspond to energies $E=0$, $E=\pm3J$, $E=\pm7J$. }
		\label{fig:12}
\end{figure}

We change the parameter  $\phi/2=Jt/\hbar$  from $-8\pi$ to $8\pi$ with step $\pi/24$ and detect the mean values $\langle\sigma_0^x(t)\sigma_2^x(t)\sigma_{13}^x(t)\rangle$  for $t=n\tau$, $\tau=\hbar\pi/J24$, $n=-N,-N+1,...N-1,N$, N=192 on quantum computer  $\textrm{ibmq\_melbourne}$ and on quantum simulator $\textrm{ibmq\_qasm\_simulator}$.    On the basis of the obtained results we find $A(\omega)$ (\ref{aw0}) (see Fig. \ref{fig:12}). The peaks of the real part of the function $A(\omega)$ at $\omega=0$, $\omega=\pm 6$,  $\omega=\pm 14$ correspond to energies $E=0$, $E=\pm3J$, $E=\pm7J$ in the system  (\ref{hams5}). Note that the peaks obtained on the basis of calculations on the quantum device (see Fig. \ref{fig:12} (b)) are not so clear as that obtained on quantum simulator $\textrm{ibmq\_qasm\_simulator}$  and in the case of spin in magnetic field, spin chain, Ising model on squared latice  (\ref{hams3}),  (\ref{hams4}). This is because the quantum protocol  Fig. \ref{fig:11} contains more gates and measurements than that considered in the previous examples (Fig. \ref{fig:1}, (Fig. \ref{fig:6}, (Fig. \ref{fig:7}) that leads to accumulation of errors. Nevertheless even in this case the method of detecting of the energy levels by studies of evolution of mean value gives as possibility to detect the energy levels.

Note that it  is not a trivial problem to find minimal or maximal eigenvalue of Ising model with spatial anisotropy. The proposed algorithm allows to solve this problem on quantum computer and we hope that using quantum computers with largest numbers qubits it will be possible to achieve quantum supremacy.

\section{Conclusions}
 The method for detecting transition energies  by studies of evolution of mean value of physical quantity on quantum computer has been proposed. In the case when the operator of a physical quantity anticommutes with the Hamiltonian of a physical system, the proposed method gives a possibility to detect energy levels of the system.
A spin in magnetic field, spin chain with Ising interaction,  Ising model on squared latice with isotropic and spatially anisotropic Ising interaction have been studied. We have examined evolution of mean values of operators anticommuting with Hamiltonians of the systems on IBM's quantum computers $\textrm{ibmq\_manila}$, $\textrm{ibmq\_melbourne}$ using quantum protocols presented on Figs. \ref{fig:1}, \ref{fig:6}, \ref{fig:8}, \ref{fig:11}  and detect energy levels of the corresponding systems (see Figs. \ref{fig:5}, \ref{fig:7}, \ref{fig:9}, \ref{fig:12}).

Note that for the case when different $J_{ij}$ takes different values with different signs there is not trivial problem to find minimal or maximal eigenvalue of Ising model (\ref{IZ}) one has
a combinatorial optimization problem. Therefore the proposed in this paper method of finding the energy levels on a quantum computer opens a possibility to achieve  quantum supremacy with development of quantum computers with largest number of qubits.

\section*{Acknowledgments}
This work was supported by Project 2020.02/0196 (No. 0120U104801) from National Research Foundation of Ukraine.

\end{document}